\documentclass{article}
\usepackage[utf8]{inputenc}
\usepackage{kotex}

\title{C-ITS 환경 및 공격 모델링}
\author{고려대학교 정보보호대학원 해킹대응기술연구실 \\ (최재웅, 송민근, 이효선, 사공채연, 박상범, 이재성, 유정도, 김휘강)}
\date{November 2023}

\usepackage[numbers]{natbib}
\usepackage{graphicx}
\usepackage{amsmath}
\usepackage{amssymb}
\usepackage{verbatim}
\usepackage{cleveref}
\usepackage{subfig}
\usepackage{titlesec}
\usepackage{makecell}
\usepackage{multirow}
\usepackage{multicol}
\usepackage{booktabs}
\usepackage{adjustbox}
\usepackage{url}

\titleformat{\paragraph}
{\normalfont\normalsize\bfseries}{\theparagraph}{1em}{}
\titlespacing*{\paragraph}
{0pt}{3.25ex plus 1ex minus .2ex}{1.5ex plus .2ex}

\begin{document}

\maketitle

\begin{abstract}
기술이 발전하면서 많은 데이터를 처리할 수 있게 되고, 도시 내 여러 요소가 다양화되고 복잡해지면서 도시들이 점차 지능화, 즉 스마트 시티로 진화하고 있다.
스마트 시티의 핵심 시스템 중 하나로 Cooperative-Intelligent Transport Systems (C-ITS)가 있다. 
C-ITS는 차량이 주행 중 운전자에게 주변 교통상황과 급정거, 낙하물 등의 사고 위험 정보를 노변 기지국을 통하여 실시간으로 제공하는 시스템이며, 도로 인프라, C-ITS 센터, 차량 단말기로 구성된다.
한편, 스마트 시티는 도시의 많은 요소들이 네트워크로 연결되고 전자적으로 제어되기 때문에 사이버 보안 문제가 발생할 수 있다.
C-ITS에서 사이버 보안 문제가 발생할 경우에는 안전 문제가 발생할 위험이 크다.
본 기술 문서는 C-ITS 환경과 제공되는 서비스를 모델링하는 것으로 스마트 시티 환경에서 보안 사고가 발생할 수 있는 Attack Surface를 식별하는 것을 목적으로 한다. 이후 식별된 Attack Surface를 기반으로 모델링된 모델 위에서 공격 시나리오와 각 단계를 구성하는 것을 목적으로 한다.
C-ITS의 개념에 대해 기술한 뒤, 우리가 정의한 C-ITS 환경 모델, 서비스 모델과 공격 시나리오 모델을 기술한다.
\end{abstract}


\subsubsection*{Acknowledgment}
This work was supported by Institute of Information \& Communications Technology Planning \& Evaluation (IITP) grant funded by the Korea government (MSIT) (No. 2021-0-00624, Development of Intelligence Cyber Attack and Defense Analysis Framework for Increasing Security Level of C-ITS)

\section{서론}
\begin{figure}[ht!]
    \centering
    \includegraphics[width=13cm,height=8cm]{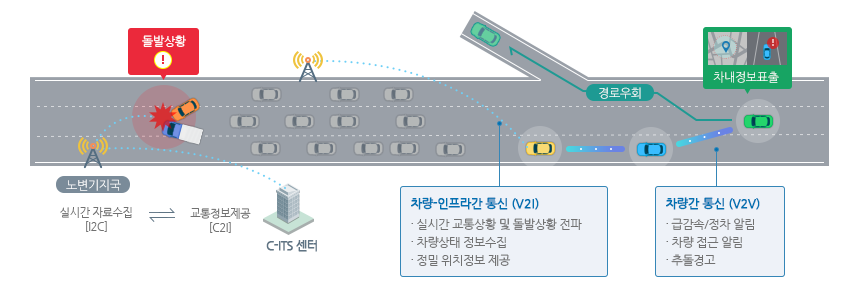}
    \caption{C-ITS Infrastructure.}
    \label{fig_cits}
\end{figure}
스마트 시티는 다양한 데이터를 수집하고, 이 데이터를 도시 운영에 사용하는 도시이다. 기술이 발전하면서 많은 데이터를 처리할 수 있게 되고, 도시 내 여러 요소가 다양화되고 복잡해지면서 도시들이 지능화, 즉 스마트 시티로 진화하고 있다. 스마트 시티의 핵심 시스템 중 하나로 Cooperative-Intelligent Transport Systems (C-ITS)가 있다. C-ITS는 교통 체계를 제어하며, 이를 위해 도시 내 차량과 교통 인프라 정보를 이용한다. 도시에서의 인적, 물적 교류는 매우 중요한 문제이기 때문에 이를 효율적으로 제어할 수 있는 C-ITS의 필요성은 매우 높다. C-ITS의 구조는 Figure \ref{fig_cits}와 같다.

안전한 스마트 시티를 구현하기 위해서는 C-ITS에 대한 보안 조치가 필수 불가결하다. 이러한 상황에서 보안 조치를 취하기 위해선 해당 시스템에 대한 위협을 식별하고, 공격 경로를 예측하여 해당하는 공격 경로에 대해 자원의 우선 순위에 따른 보안 수준을 구성할 필요가 있다. 다시 말해 스마트 시티 환경에서 사이버 보안 문제를 예방하고 최소화 하기 위해서는 C-ITS 보안 대책이 필요하다.

본 기술 문서에서는 C-ITS 보안을 위해 C-ITS 환경을 구체적으로 모델링하고, 모델링 된 C-ITS 환경에 대해 공격 시나리오를 기술하는 것을 목적으로 한다. Section \ref{sec:background}에서는 환경 모델링에 필요한 배경 지식에 대해 기술한다. Section \ref{sec:env-model}, \ref{sec:service-model}에서는 본 기술 문서에 기술된 C-ITS 환경 모델을 구성하기 위해 사용된 모델링 방법에 대해 기술한다. Section \ref{sec:attack-model}에서는 모델링 된 C-ITS 환경을 기반으로 공격 시나리오를 기술한다.

\section{배경 및 용어}
\label{sec:background}
\subsection{C-ITS}

\subsubsection{C-ITS 개념}
C-ITS는 차량간 통신을 개선하기 위해 제안된 개념으로서 도로 안전 증가, 교통관리 최적화 등 목적을 가진다. 해당 목적을 달성하기 위해서 차량간 (Vehicle-to-vehicle), 차량과 도로 인프가간 (Vehicle-to-infrastructure), 차량과 보행자간 (Vehicle-to-pedestrian) 등 통신 서비스를 통해서 주고받는 서비스를 제공한다.

\subsubsection{C-ITS 필요성}
C-ITS의 필요성은 도로 교통의 복잡성의 증가로부터 비롯된다. 도로 위의 차량 수의 증가로 인해 교통 체증과 교통사고 발생률이 상승하고 있으며, 이로 인한 경제적 손실또한 발생한다. C-ITS는 이러한 변화에 맞춰 교통 흐름을 최적화하고, 안전한 운행을 지원하여 위에서 기술한 문제들을 해결하기 위한 필수적인 기술로 간주된다. 추가적으로 친환경 교통 정책에 부합하도록 효율적인 에너지 관리 및 배출 감소를 위한 기술적 지원을 제공함으로써 환경 지속 가능성을 강화한다.

\paragraph{교통사고 예방을 통한 안전성과 이동성 향상}
C-ITS 환경에서의 실시간 데이터 교환을 통해 차량의 위치, 속도, 및 주변 환경에 대한 정보를 공유하여 사고를 사전에 감지하고 회피한다. 또한, 교통 신호 조절 및 교차로 관리를 최적화하여 교통흐름을 조절해서 교통사고 발생 가능성을 줄인다. 이동성 측면에서는 실시간 교통 정보를 운전자에게 제공함으로써 도로 체증을 회피하고 최적의 경로를 찾을 수 있게 한다. 이러한 방법들로 C-ITS는 교통 안전성과 이동성 향상에 중요한 역할을 한다.

\subsubsection{RSU}
RSU (Roadside Unit)은 C-ITS의 구성요소로서 차량 간 통신 및 차량과 도로 인프라 간 통신을 지원한다. RSU는 차량에게 도로 상태 정보, 교통 신호, 사고 경고 및 다양한 서비스와 관련된 데이터를 제공하고 차량과 중앙 서버 또는 RSU Cloud와 연결되어 효율적인 데이터 교환을 촉진하며 교통 흐름을 최적화하는 데 기여한다. 추가적으로 RSU는 또한 향상된 신호 제어 및 교통 관리를 통해 도로 네트워크의 효율성을 향상시키고, 교통 안전성을 증진시키는 데 중요한 역할을 한다.

\paragraph{RSU Cloud}
RSU Cloud는 여러 RSU에서 수집된 데이터를 통합하고 처리하는 클라우드 기반 시스템이다. 주로 교통 관리, 도로 안전 및 다양한 C-ITS 서비스에 대한 중앙화된 데이터 처리하는 작업을 진행한다. 대량의 데이터를 분석하여 도로 네트워크의 상태를 모니터링하고, 효율적인 교통 관리 결정을 내리는 데 도움을 주고 향상된 서비스 및 교통 관리를 위해 RSU와 차량 간의 실시간 데이터 교환을 지원하며, 빅데이터 기술을 활용하여 도로 네트워크의 전반적인 효율성을 향상시킨다.

\subsubsection{OBU}
OBU (On-Board Unit)은 차량 내 장치로 차량 간 통신, 차량과 도로 인프라 간 통신을 위한 인터페이스를 제공하기도하며 RSU와의 통신을 통해 도로 정보, 교통 신호, 사고 경고 등의 데이터를 수신하고 송신한다. OBU는 차량 운전자에게 경고 및 안전 서비스를 제공하고, 동시에 도로 네트워크에 차량의 현재 상태를 보고하는 데 사용된다. 

\subsubsection{PDU}
\begin{figure}[!ht]
\centering
    \includegraphics[width=12.0cm]{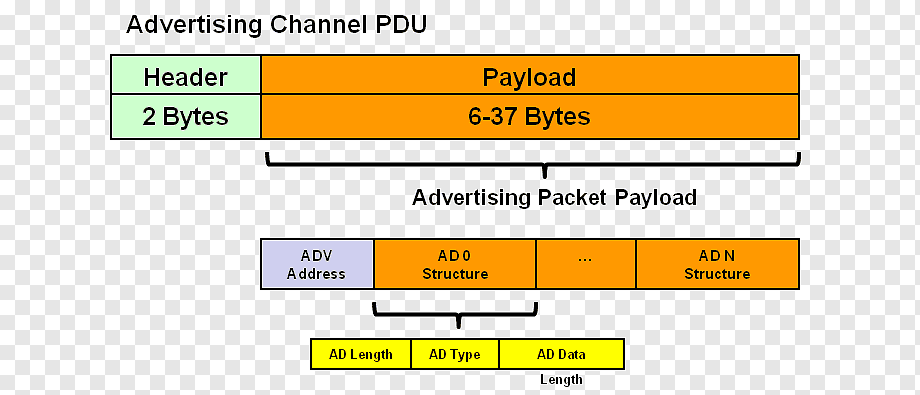}
\caption{Protocol Data Unit.}
\label{fig:PDU}
\end{figure}
PDU (Protocol Data Unit)은 통신 프로토콜에서 데이터를 교환하는 데 사용되는 패킷 형식의 유닛이다. 이는 데이터를 특정 프로토콜의 표준 형식으로 구성하여 안전하고 효율적인 전송을 가능케 하며, 데이터의 일관성을 유지하고 통신의 신뢰성을 향상시킨다. 주로 차량 간 통신 및 차량과 도로 인프라 간 통신에서 사용되며, 정보를 패킷 형태로 분할하여 전송하고 수신하는 데 중요한 역할을 한다. PDU는 다양한 C-ITS 서비스에서 데이터의 일관성과 정확성을 보장하여 안전한 차량 통신을 지원한다. PDU는 Figure \ref{fig:PDU}와 같다.

\section{환경 모델링}
\label{sec:env-model}
\subsection{소개}
모델링이란, C-ITS 환경에서 Road-Side Unit(RSU) \& RSU Cloud, On-Board Unit(OBU), Central Cloud에 대한 시스템 요구사항과 서비스 요구사항을 Depth에 따라 간결하게 표현한 것을 의미한다. 본 기술문서에서 모델링 한 시스템 및 서비스는 \textit{depth-0}, \textit{depth-1}, \textit{depth-2}로 순차적으로 표현된다. 각 요소에 대한 시스템 요구사항은 독일의 Dresden에 구성된 Backend Service와 RSU를 참조하여 모델링하며\cite{strobl2019c}, OBU에 대해 On-Board Equipment와 통신까지 상세 모델링 수행한다\cite{9621229}. 본 장에서 기술하는 RSU Cloud와 Central Cloud는 Cloud 서비스를 활용하여 각 요소간의 통신을 수행하며 각각 RSU 혹은 전체적인 C-ITS 환경을 제어한다.

\subsection{환경 시스템 구성}
\subsubsection{Depth-0 Modeling}

\begin{figure}[!ht]
\centering
    \includegraphics[width=12.0cm]{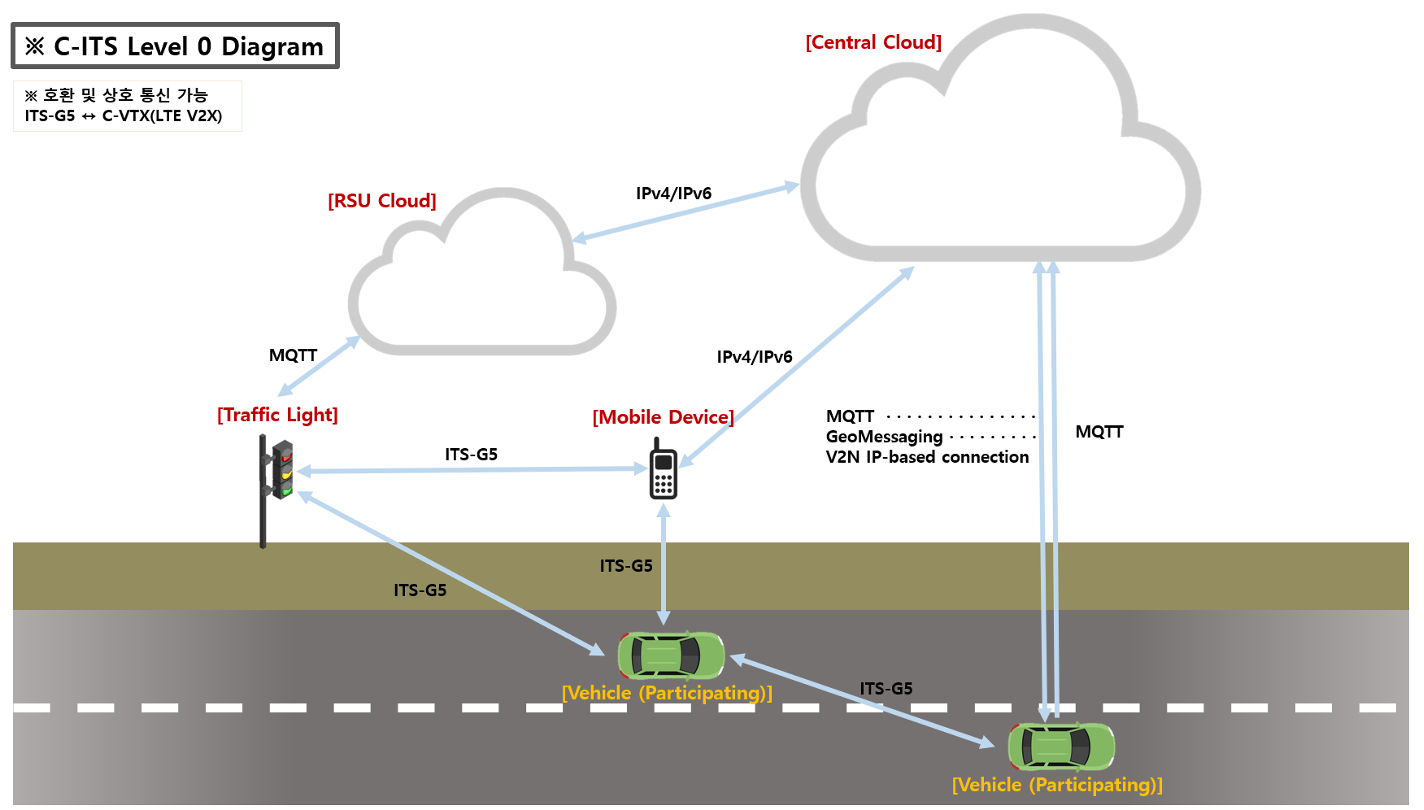}
\caption{C-ITS Communication Overview.}
\label{fig:LZ}
\end{figure}

Depth-0 Modeling은 각 요소간 통신 사이에 필요로 하는 통신 프로토콜을 Figure \ref{fig:LZ}와 같이 전체적으로 표현한다. Traffic Light는 RSU가 될 수 있는 요소 중 하나이며, RSU는 MQTT 프로토콜을 활용해 RSU Cloud와 통신한다. 이를 통해 RSU가 수집한 데이터를 RSU Cloud와 Central Cloud에 저장하게 된다. Cloud System간의 통신은 기존과 동일하게 인터넷 프로토콜을 활용하며, RSU와 Mobile Device, RSU와 Vehicle은 ITS 시스템에서 구현된 ITS-G5 무선 통신을 수행한다.

\subsubsection{Depth-1 Modeling}

Depth-1 Modeling은 Depth-0 에서 필요로 하는 각 요소에 대한 모델링을 표현한다. 

\paragraph{Hybrid Communication}
\label{Section:HC}

\begin{figure}[!ht]
\centering
    \includegraphics[width=8.0cm]{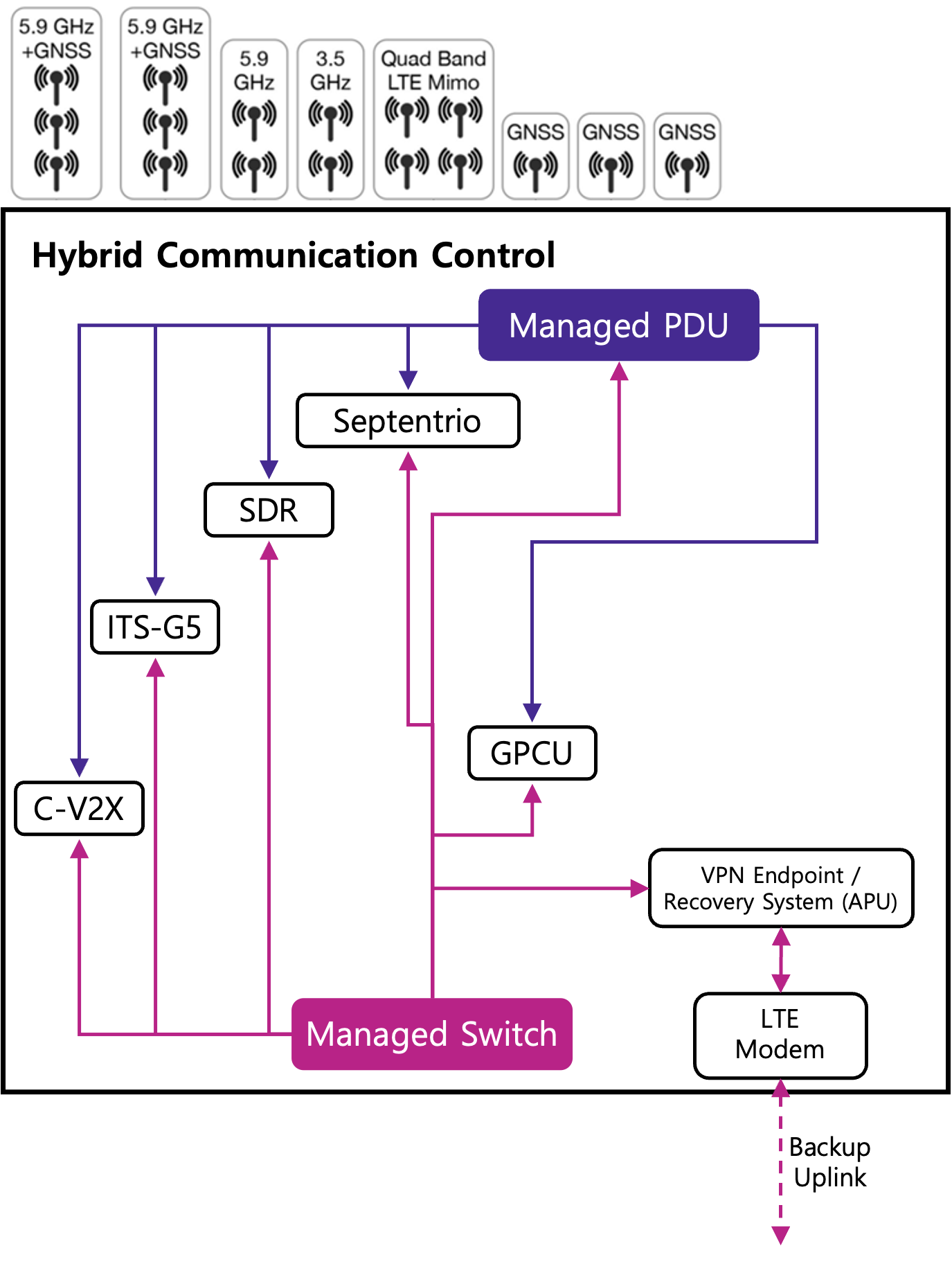}
\caption{Hybrid Communication Framework.}
\label{fig:HC}
\end{figure}

Hybrid Communication은 OBU, RSU, Central Cloud 등의 스마트 시티 내 요소들간의 통신을 수행할 때 이기종 간 통신 프로토콜을 맞추어 변환하기 위해 사용되는 통신 기법이다. 이는 둘 이상 통신 인터페이스를 통해 하나의 메시지를 병렬 처리 하는 방법이다. 본 장에서 소개하는 Hybrid Communication의 Framework는 Figure \ref{fig:HC}와 같다. 

Figure \ref{fig:HC}을 기반으로 Protocol Data Unit(PDU)을 관리하고, 필요한 통신 인터페이스에 맞추어 병렬 처리하여, Endpoint를 통해 데이터를 전송한다.

\paragraph{On-Board Unit(OBU)}

\begin{figure}[!ht]
\centering
   \includegraphics[width=8.5cm]{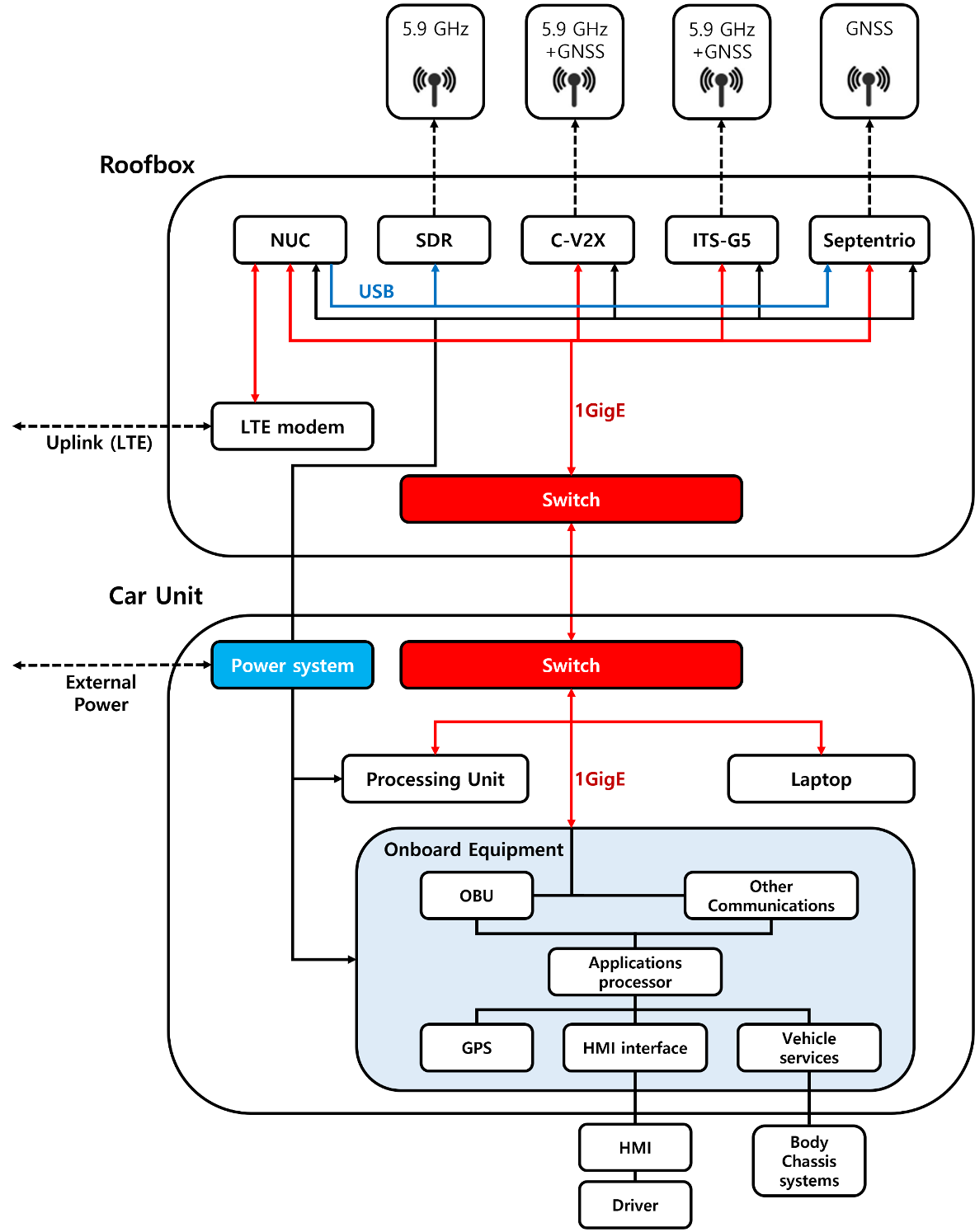}
\caption{On-Board Unit Overall Framework.}
\label{fig:OOF}
\end{figure}

OBU는 차량 혹은 운송 수단에 장착되는 전자 장치로, 운송 및 통신 기능을 제공한다. 스마트 시티에서 스마트 도로, 도로 요금 수납, 자율 주행, 교통 관리 시스템 등과 통합되며, 또 다른 OBU, 후술할 RSU \& RSU Cloud 혹은 Central Cloud와 주로 통신을 수행한다. 

본 장에서 소개하는 OBU의 전체적인 Framework는 Figure \ref{fig:OOF}와 같다. 차량과 차량 상단의 Roofbox를 활용하여 구성된 해당 Framework는 차량과 Roofbox 각각에 통신을 위한 Switch가 존재하며, 운전자와 On-Board Equipment 간에 Human Machine Interface를 제공하여 운전자가 해당 시스템의 각종 서비스를 사용할 수 있도록 한다. Roofbox는 일반적으로 다른 차량 혹은 운송 수단과의 통신, RSU \& RSU Cloud와의 통신, Central Cloud와의 통신을 위해 존재한다. 해당 통신은 송신자와 수신자에 따라 사용되는 규격이 달라지는 것을 고려하여 Hybrid Communicationd을 지원하도록 구성한다.

\paragraph{Road-Side Unit(RSU) \& RSU Cloud}

\begin{figure}[!ht]
    \centering
    \includegraphics[width=12cm]{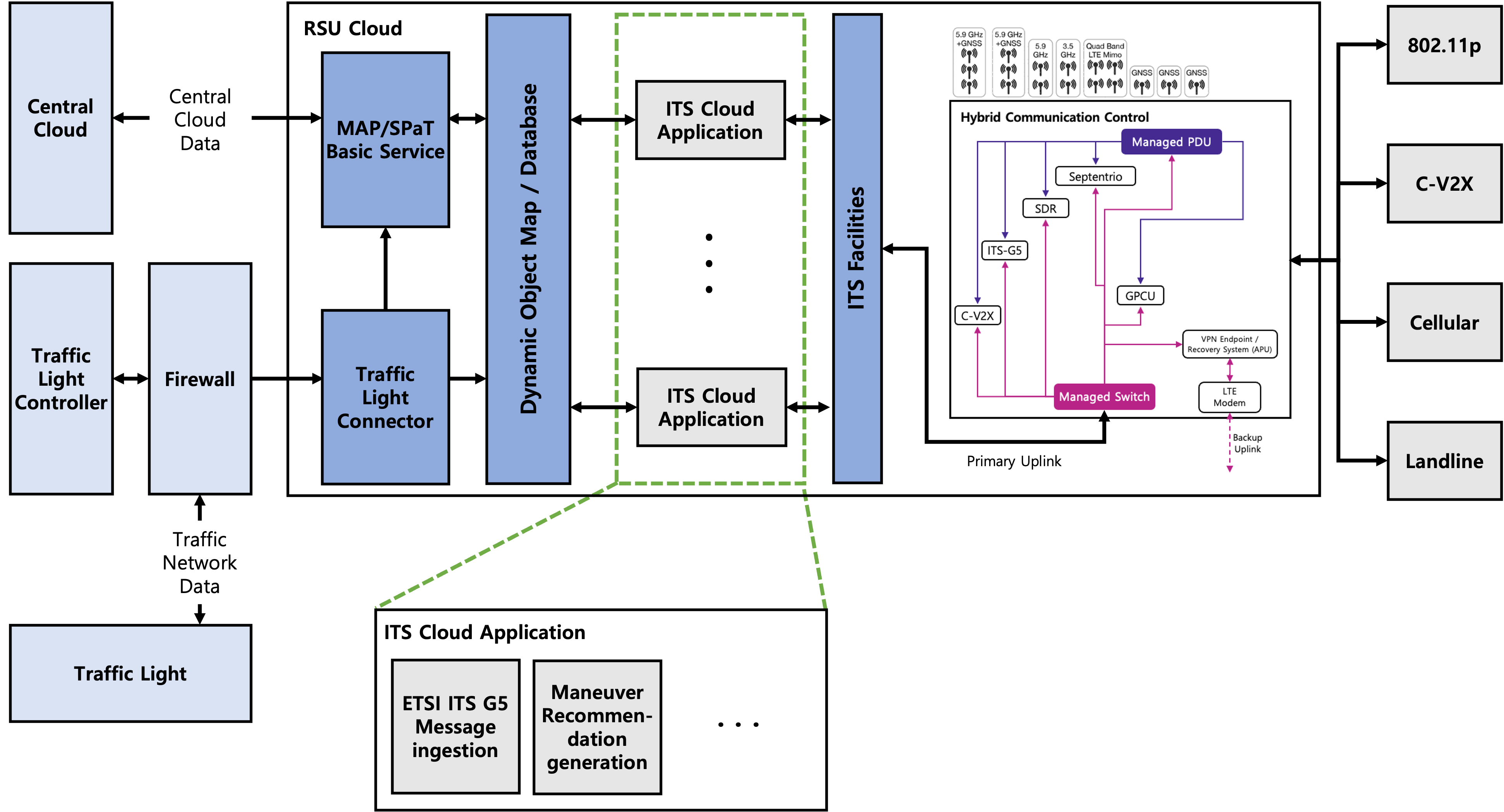}
    \caption{Road-Side Unit Cloud Overall Framework.}
    \label{fig:RCOF}
\end{figure}

RSU는 가로등, 신호등 등과 같은 도로에 존재할 수 있는 도로변 장치에 장착되는 전자 장치로, 통신 기능을 제공한다. 스마트 시티에서 Central System의 정보 혹은 직접 수집한 정보 등을 받아 연결된 다른 RSU 혹은 연결된 차량의 OBU로 데이터를 전송한다.

본 장에서 소개하는 RSU는 앞선 설명과 같이 도로변 장치에 장착되는 장치가 존재하며, 추가적으로 RSU Cloud를 구성한다. RSU Cloud의 전체적인 Framework는 Figure \ref{fig:RCOF}와 같다.

RSU Cloud에는 앞서 \ref{Section:HC}에서 설명한 Hybrid Communication을 구현하고 있으며, 이를 통해 다른 RSU 내지는 OBU와 직접 통신한다. 또한, 데이터 수집을 위해 데이터베이스를 두고 있으며, 수집한 데이터를 기반으로 ITS Cloud Application을 이용해 가공한다. 가공된 데이터는 다시 데이터베이스에 저장되거나 Hybrid Communication을 통해 Vehicle 혹은 Mobile Device에 전송될 수 있다. 또한, RSU Cloud에서 수집하고 가공한 모든 데이터는 Central Cloud로 전송되어, 필요에 따라 직접 연결되지 않은 RSU Cloud 혹은 Vehicle 등에 Central Cloud에서 2차 가공하여 데이터를 전송할 수 있도록 한다.

\paragraph{Central Cloud}

\begin{figure}[!ht]
    \centering
    \includegraphics[width=12cm]{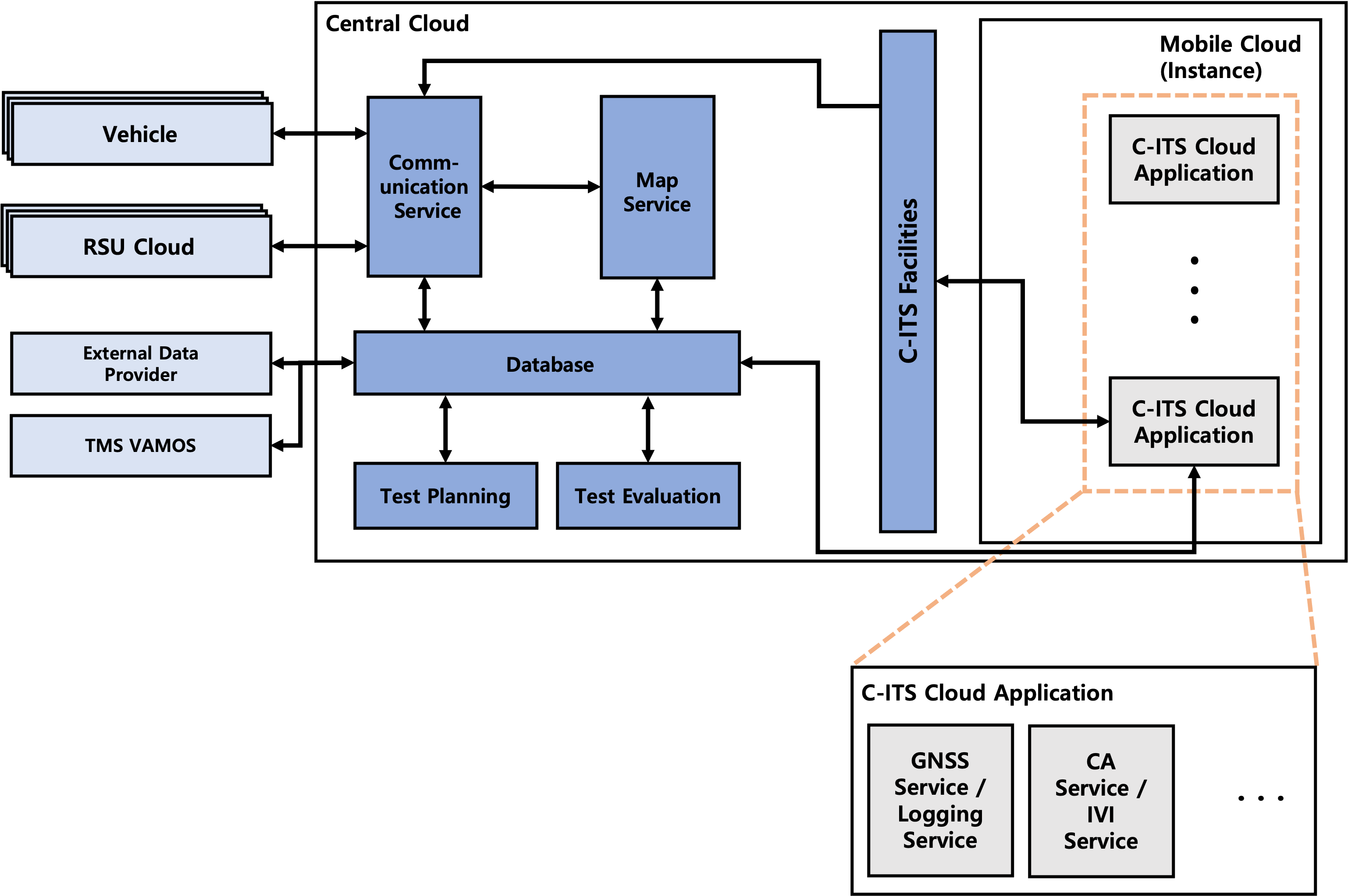}
    \caption{Central Cloud Overall Framework.}
    \label{fig:CCOF}
\end{figure}

Central Cloud는 RSU와 OBU가 수집한 대부분의 데이터가 저장되는 곳이며, 저장된 데이터를 가공하여 사용자(OBU, RSU, Mobile Device 등)가 필요한 데이터를 제공하는 Cloud 서비스이다. Central Cloud 내 Mobile Cloud는 Docker Container 혹은 AWS Instance 등으로 구현될 수 있는 Instance 이며, 해당 Instance 내에 C-ITS Cloud에서 제공 가능한 여러 Application이 존재한다. 본 장에서 소개하는 Central Cloud는 Figure \ref{fig:CCOF}와 같다.

구체적으로 Central Cloud는 외부 데이터 제공자, Vehicle, RSU Cloud 등으로 부터 데이터를 수집하게 되는데, 해당 데이터를 데이터베이스에 저장한 뒤, Mobile Cloud에 존재하는 Application Service를 이용하여 2차 가공한다. 2차 가공된 데이터는 스마트 시티의 테스팅 계획을 수립하거나 평가하는 등의 업무 또한 수행할 수 있으며, 2차 가공된 데이터는 외부 데이터 제공자, Vehicle 혹은 RSU Cloud에도 전송할 수 있다. Central Cloud의 Application은 데이터 저장을 위해 포맷을 맞추는 과정, Logging 등의 서비스가 포함될 수 있으며, 자세한 내용은 depth-2 Modeling에서 후술한다.

\subsubsection{depth-2 Modeling}

\begin{figure}[!ht]
    \centering
    \includegraphics[width=12cm]{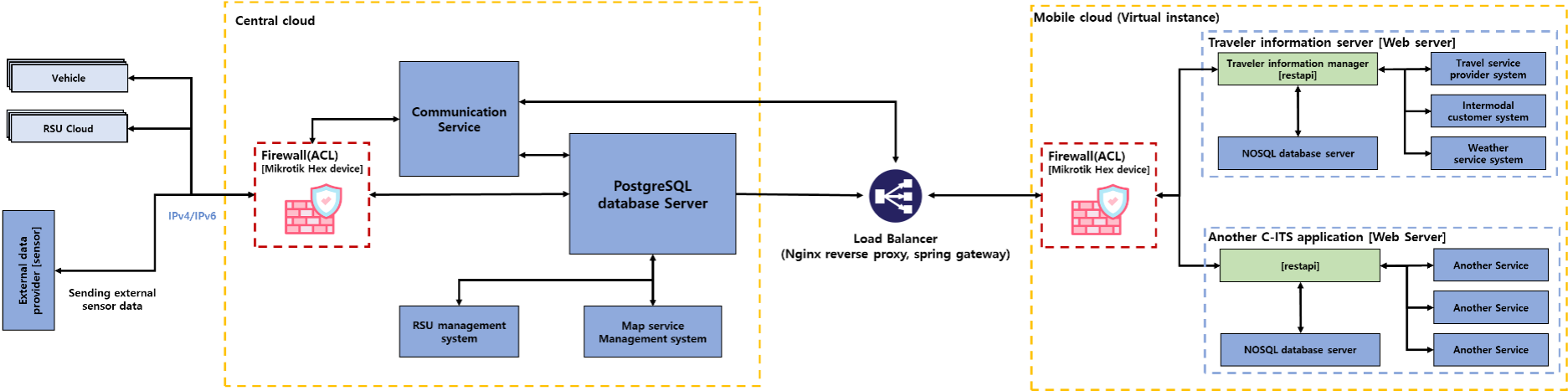}
    \caption{Central Cloud and Mobile Cloud Detail Framework.}
    \label{fig:Detail-CCOF}
\end{figure}

본 기술 문서에서 소개하는 \textit{depth-2} Modeling은 Central Cloud의 상세 모델링이다.

상세 모델링은 Figure \ref{fig:Detail-CCOF}과 같이 표현되며, 각 서비스에 대응되는 구체적이고 대체가능한 서비스를 모델링하고 있으며, 내부 Instance인 Mobile Cloud에 대한 추상화 된 서비스 객체와 구현체 모두 제공한다. Mobile Cloud는 각 서비스 별로 별도 구성되며, Central Cloud에 각 서비스를 요청할 때, 요청에 따른 Load Balancing을 수행한다. 더욱 상세하게 Mobile Cloud에는 C-ITS 환경에서 제공될 수 있는 모든 서비스가 올라갈 수 있으며, Figure \ref{fig:Detail-CCOF} 우측의 Mobile Cloud 내 하단의 파란 점선 내부의 요소가 추상화 된 서비스이다. 각 박스는 상단의 파란 점선 내부 요소들과 같이 C-ITS 환경에서 제공하고자 하는 서비스(\textit{e.g.,} 신호 관리, 데이터 웨어하우스)를 구현한다.

\section{서비스 모델링}
\label{sec:service-model}
\subsection{소개}
본 기술 문서는 모델링한 환경을 기반으로 C-ITS 환경에서 제공되는 각종 서비스를 모델링하여 제공한다. 해당 서비스는 미국의 Architecture Reference for Cooperative and Intelligent Transportation(ARC-IT)를 참조하여 모델링된 시스템에 적용한다\cite{ARC-IT}.

\subsection{서비스 구성}
\subsubsection{Parking Management 01: Parking Space Management}
\begin{figure}[!ht]
    \centering
    \includegraphics[width=12cm]{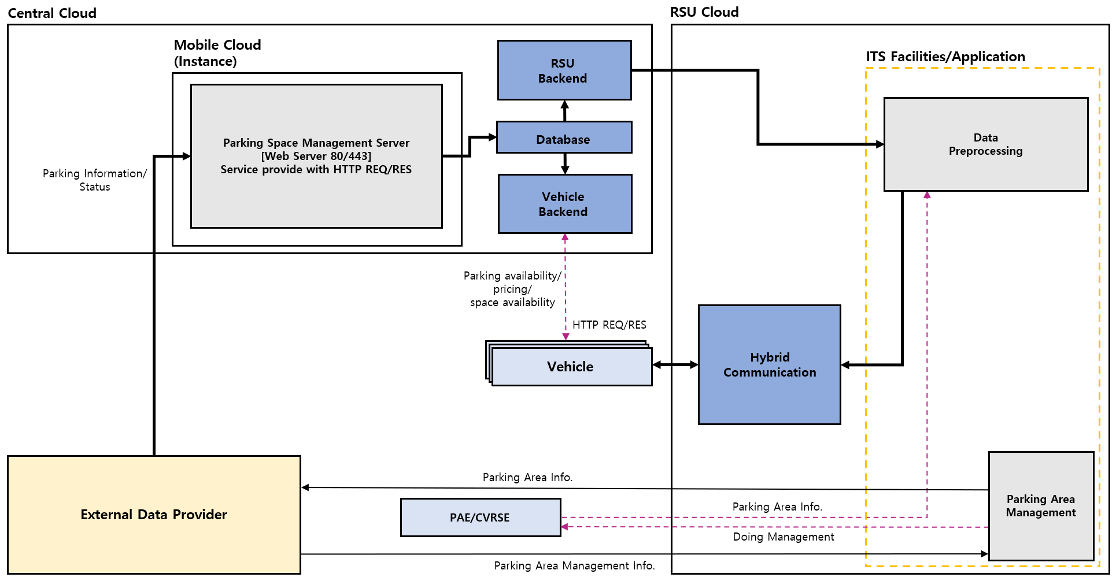}
    \caption{Parking Space Management Service Diagram.}
    \label{fig:PM01}
\end{figure}

Parking Management 01은 주차장 차고 및 기타 주차 공간 및 시설의 주차 공간을 모니터링하고 관리하는 서비스 패키지이다. 주차 공간 점유율 및 가용성을 모니터링하여 주차 운영 관리를 지원하는 C-ITS 서비스로 Vehicle(주차 공간 사용자)이 Central Cloud로 특정 주차장에 대해 주차 공간 정보 및 가격 등의 원하는 정보를 요청하면 응답받게 된다. 주차 공간의 정보는 지속적으로 RSU와 RSU Cloud를 통해 외부 데이터 제공자에게 수집되고 주차 정보와 상태를 외부 데이터 제공자가 Central Cloud의 Instance에 지속적으로 업데이트한다.

\subsubsection{Public Safety 03: Emergency Vehicle Preemption}
\begin{figure}[!ht]
    \centering
    \includegraphics[width=12cm]{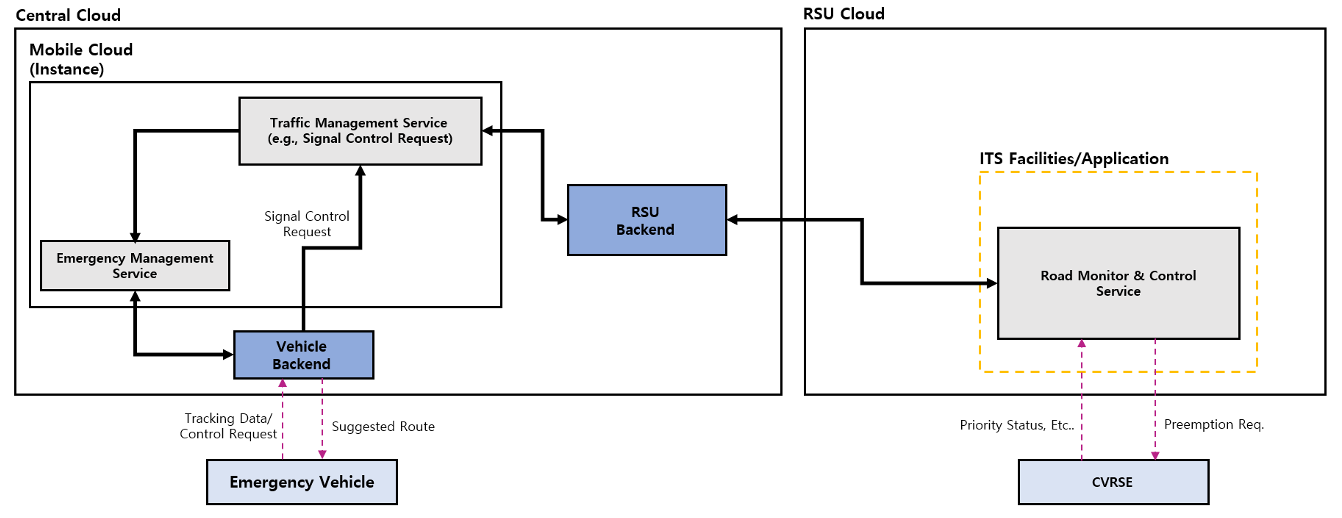}
    \caption{Emergency Vehicle Preemption Service Diagram.}
    \label{fig:PS03}
\end{figure}

Public Safety 03은 공공 안전 및 응급 구조 차량의 신호 선점 기능을 제공하기 위한 서비스 패키지이다. 등록된 응급 차량이 Central Cloud와 통신하여 자신의 위치 정보 및 신호 제어 요청을 전송한다. 이에 따라 Central Cloud는 응급 차량에 경로를 전송하며, 해당 경로에 따라 RSU Cloud가 Central Cloud로부터 받은 요청을 기반으로 신호 변경을 수행한다.

\subsubsection{Support 01: Connected Vehicle System Monitoring and Management}
\begin{figure}[!ht]
    \centering
    \includegraphics[width=12cm]{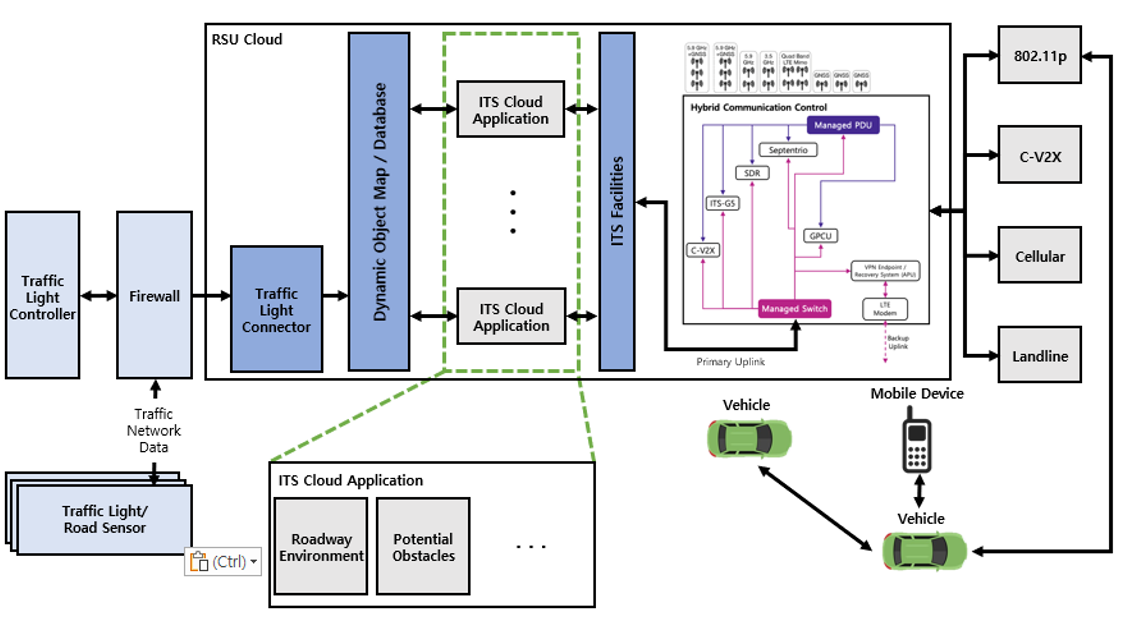}
    \caption{Connected Vehicle System Monitoring and Management Service Diagram.}
    \label{fig:SU01}
\end{figure}

Support 01은 각종 애플리케이션 혹은 장치(RSU, Vehicle, Mobile Device 등)에 필요한 모니터링, 관리 및 제어 서비스를 제공하는 서비스 패키지이다. 특정 Vehicle을 기준으로 근처 다른 차량 및 주변 환경과 RSU, 사람이 휴대한 Mobile Device 등으로부터 필요로 하는 정보를 받아 업데이트를 수행한다.

\subsubsection{Traveler Information 03: Dynamic Route Guidance}
\begin{figure}[!ht]
    \centering
    \includegraphics[width=12cm]{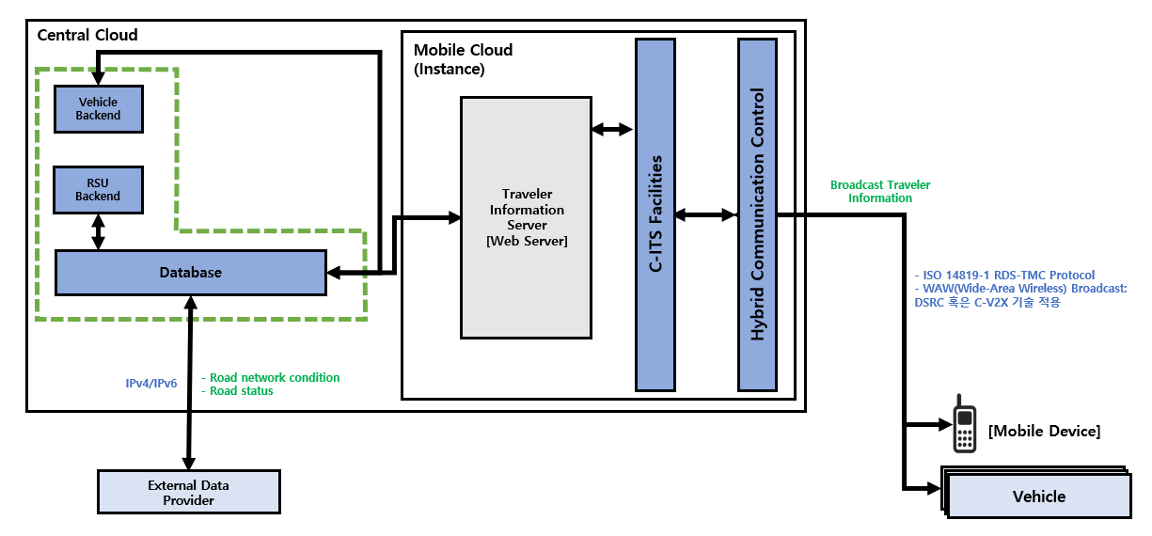}
    \caption{Dynamic Route Guidance Service Diagram.}
    \label{fig:TI03}
\end{figure}

Traveler Information 03은 특정 상황에 대응하는 경로를 계획하고 안내를 제공하는 서비스 패키지이다. 외부 데이터 제공자 혹은 자체적으로 수집하여 데이터베이스에 보유하고 있는 교통 및 사고 정보 등을 기반으로 여러 Vehicle, 사람의 Mobile Device 등에 원활히 이동이 가능한 경로를 설정하여 안내를 제공한다.

\section{공격 모델링}
\label{sec:attack-model}
\subsection{소개}
본 기술 문서는 모델링한 서비스를 기반으로 서비스에 대한 공격 시나리오를 구성하여 제공한다. 본 장에서는 각 요소를 기반으로 이루어지는 공격 시나리오를 구성하며, 각 공격 단계에 따른 CVE 취약점을 제공한다. CVE 취약점은 CVE 데이터베이스\cite{CVE}에서 수집한 취약점이다.

\subsection{시나리오 구성}

\subsubsection{Attack Scenario 1}
\begin{figure}[!ht]
    \centering
    \includegraphics[width=12cm]{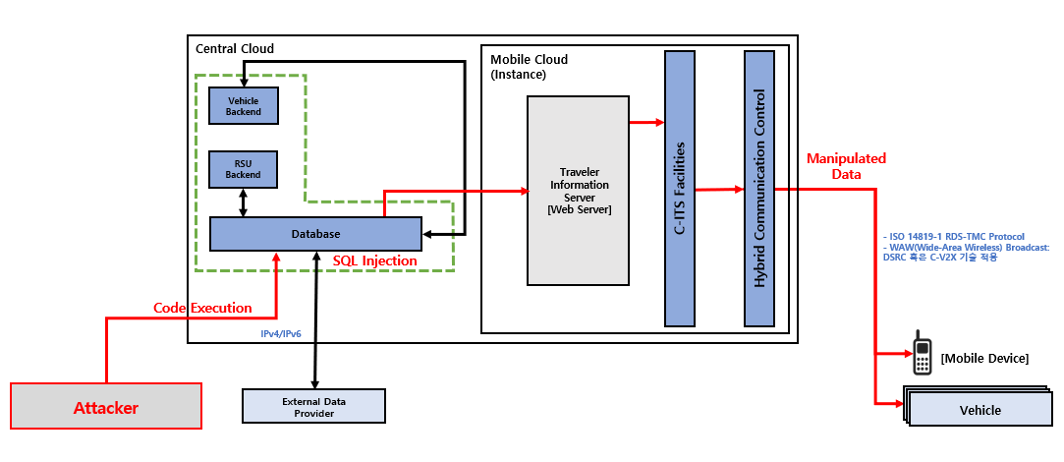}
    \caption{Attack Scenario 1 Diagram about Central Cloud to OBU or Mobile Device.}
    \label{fig:AS1}
\end{figure}
본 공격 시나리오 1은 원격지의 공격자가 Central Cloud에 대한 외부 데이터 제공자로 가장하는 것으로 이루어지는 공격이다. 공격자는 IPv6 통신에 외부 데이터 제공자로 참여하게 되고, 이를 통해 악의적인 원격 코드를 실행하여 데이터베이스의 정보를 조작한다. 조작된 데이터는 Vehicles와 Mobile Devices에 전달되기에 본 시나리오는 이를 악용한 잘못된 정보 전달로 인해 발생되는 무결성 침해 및 무차별 공격에 대한 시나리오이다.
본 공격 시나리오를 수행하기 위한 시스템 요구사항은 다음과 같다:
\begin{itemize}
    \item IPv6 in Central Cloud
    \item MySQL Database in Central Cloud
\end{itemize}
다음 하위 Section 부터 본 공격 시나리오에 대한 공격 단계를 CVE 취약점에 대응하여 설명한다.

\paragraph{Attack Step 1}
\begin{itemize}
    \item 악용된 취약점
    \begin{itemize}
        \item \textbf{CVE-2020-27338} \cite{CVE-2020-27338}
    \end{itemize}
\end{itemize}
공격 단계 1에서 공격자는 DHCPv6 Client 구성 요소의 부적절한 입력 유효성 검사를 악용하여 침투한 뒤 악의적인 코드를 실행할 수 있도록 준비한다. 해당 취약점은 6.0.1.68 이전의 Treck IPv6에서 발생한 문제이다. 원격 공격자가 범위를 벗어난 Read를 유발하고 인접한 네트워크 접근을 통해 DoS를 유발할 수 있다. 이를 응용하여 인접한 네트워크에 접근을 수행한다.

\paragraph{Attack Step 2}
\begin{itemize}
    \item 악용된 취약점
    \begin{itemize}
        \item \textbf{CVE-2022-30927} \cite{CVE-2022-30927}
    \end{itemize}
\end{itemize}
공격 단계 2에서 공격자는 취약점을 활용하여 취약한 \textit{id} 매개변수를 활용하여 SQL 명령을 실행한다. 해당 취약점은 취약점을 악용할 수 있는 지점까지 접근해야 하며, 공격 단계 1을 통해 접근된 시점에 악용 가능하다. 악용되는 \textit{id} 매개변수는 MySQL이 데이터베이스로 사용될 때, Simple Task Scheduling System 1.0에 존재하는 취약점 요소이다. 이를 통해 공격자는 데이터베이스 내의 정보를 위변조하여 데이터의 무결성을 해친다.

\paragraph{Attack Step 3}
공격 단계 1과 2가 모두 성공한 경우 Database는 잘못된 정보로 조작되어 있게 된다. 이로 인해 조작된 도로 정보와 도로 상태, 신호 정보 및 차량 상태 등의 정보가 잘못되어 Vehicle의 이동 경로를 잘못 안내하거나 최악의 경우 교통 사고로 이어질 수 있는 위험한 사고를 야기할 수 있다.

\subsubsection{Attack Scenario 2}
본 공격 시나리오 2는 원격지의 공격자가 자신의 Device를 기반으로 악의적인 Traffic Signal을 특정 RSU에 전송하는 것으로 이루어지는 공격 시나리오이다. 이를 통해 Mobile Cloud의 서비스를 장악하고, 장악한 서비스를 기반으로 각종 공격을 수행한다.
본 공격 시나리오를 수행하기 위한 시스템 요구사항은 다음과 같다:
\begin{itemize}
    \item MQTT in Vehicle Backend
    \item Spring Framework \> 5.2.15 version \& \> 5.3.7 version in Mobile Cloud Service
    \item SNMPv3 from Mobile Cloud Service to RSU Backend
\end{itemize}
다음 하위 Section 부터 본 공격 시나리오에 대한 공격 단계를 CVE 취약점에 대응하여 설명한다.

\begin{figure}[!ht]
    \centering
    \includegraphics[width=12cm]{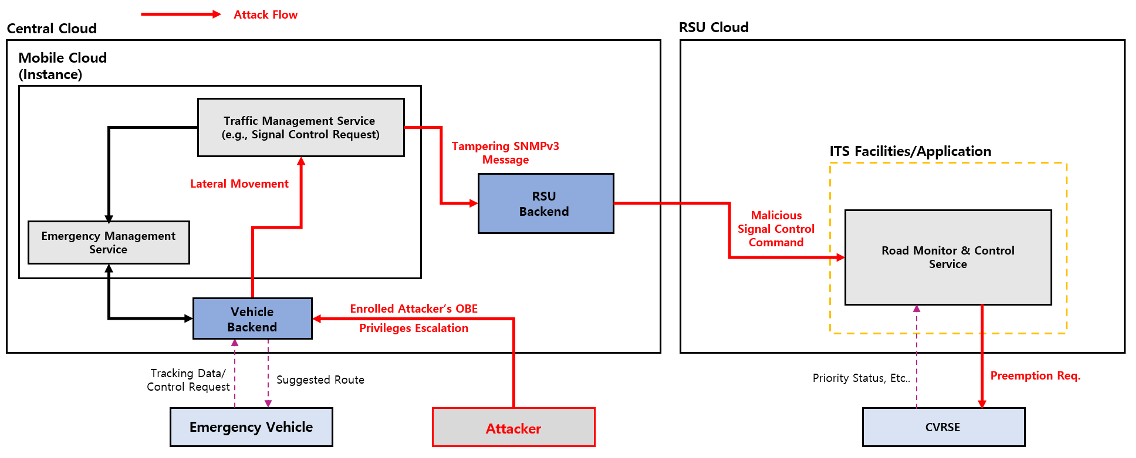}
    \caption{Attack Scenario 2 Diagram about Central Cloud to CVRSE through RSU Cloud.}
    \label{fig:AS2}
\end{figure}

\paragraph{Attack Step 1}
\begin{itemize}
    \item 악용된 취약점
    \begin{itemize}
        \item \textbf{CVE-2019-5432} \cite{CVE-2019-5432}
    \end{itemize}
\end{itemize}
공격 단계 1에서 공격자는 취약점을 통해 device에서 Mobile Cloud에 접근하며 침입을 수행한다. 이는 MQTT Broker가 잘못 구성되어 발생하는 Buffer Over-read를 악용한 공격이며 이를 통해 공격자는 자신의 device를 Mobile Cloud에 응급 차량으로 등록을 요청한다.

\paragraph{Attack Step 2}
\begin{itemize}
    \item 악용된 취약점
    \begin{itemize}
        \item \textbf{CVE-2021-22118} \cite{CVE-2021-22118}
    \end{itemize}
\end{itemize}
공격 단계 2에서 공격자는 응급 차량으로 등록된 자신의 Device를 기반으로 침입에 성공한 뒤, Spring Boot 내 WebFlux 취약점을 이용하여 권한 상승을 수행한다. 해당 취약점을 기반으로 Mobile Cloud 내 존재하는 Traffic Management Service에 대해 측면 이동 공격을 수행함과 동시에 권한 상승을 수행한다.

\paragraph{Attack Step 3}
\begin{itemize}
    \item 악용된 취약점
    \begin{itemize}
        \item \textbf{CVE-2022-43870} \cite{CVE-2022-43870}
    \end{itemize}
\end{itemize}
공격 단계 3에서 공격자는 획득한 권한을 기반으로 악성 신호 제어 명령 메시지를 전송하여 신호 변경을 방해 하는 등의 사고 유발 공격을 수행한다. 취약점을 악용하여 SNMPv3 메시지를 변조하여 Credentials를 획득하는 것을 통해 해당 서비스의 메시지를 변조할 수 있다.

\bibliographystyle{plain}
\bibliography{main}

\begin{figure}[b] %%% 이건 1행 2열짜리 그림 넣는 방법(즉, 가로로 2개 그림 넣기)
\centering
\subfloat{
\includegraphics[width=0.35\linewidth]{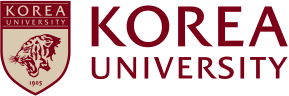}
}
\centering
\subfloat{
\includegraphics[width=0.35\linewidth]{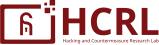}
}

\end{figure}

\end{document}